\documentclass[conference]{IEEEtran}
\IEEEoverridecommandlockouts
\usepackage{cite}
\usepackage{amsmath,amssymb,amsfonts}
\usepackage{algorithmic}
\usepackage{graphicx}
\usepackage{textcomp}
\usepackage{xcolor}
\usepackage{array}
\usepackage{cite}
\pdfoutput=1

\def\BibTeX{{\rm B\kern-.05em{\sc i\kern-.025em b}\kern-.08em
    T\kern-.1667em\lower.7ex\hbox{E}\kern-.125emX}}
\begin{document}

\title{On Tethered Satellite System Current Enhancement}

\author{\IEEEauthorblockN{Chunpei Cai, Shiying Cai}
\IEEEauthorblockA{\textit{Dept. of Mechanical Engineering-Engineering Mechanics} \\
\textit{Michigan Technological University}\\
Houghton, MI 49931, USA \\
David L. Cooke \\
Space Vehicle Directorate, AFRL, Kirtland Air Force Base \\
2000 Wyoming Blvd SE, Albuquerque, NM 87123 \\
Email:ccai@mtu.edu}
}

\maketitle
\thispagestyle{plain}
\pagestyle{plain}

\begin{abstract}
Improvements of investigations on the Tethered Satellite System (TSS)-1R electron current enhancement due to magnetic limited collections are reported. New analytical expressions are obtained for the potential and temperature  changes across the pre-sheath.  The  mathematical treatments in this work are more rigorous than one past approach. More experimental measurements collected in the ionosphere during the TSS-1R mission are adopted for validations. The relations developed in this work offer two bounding curves for these data points quite successfully; the average of these two curves is close to the curve-fitting results for the measurements; and an average of 2.95 times larger than the Parker-Murphy theory is revealed.  The results indicate that including the pre-sheath analysis is important to compute the electron current enhancement due to magnetic limitations.
\end{abstract}

\begin{IEEEkeywords}
space tether system; Parker-Murphy theory; current enhancement; pre-sheath model
\end{IEEEkeywords}

\section{Introduction}
Tethered Satellite Systems (TSS) can be deployed for a variety of important applications, including~electro-dynamic propulsion, momentum exchange, tidal stabilization,  artificial gravity, deployment of sensors or antennas, and orbital plasma dynamics.  They are helpful to investigate characterizations of  the system  current--voltage ({I--V})  response in the ionosphere plasma environment, the characterization of the physics of the satellite sheath, or current collection, and of overall current closure. The orbital motion through the Earth's geomagnetic field allows orbiting tethered systems to generate a much higher electromotive force.  Especially,  spacecraft charging is a serious and practical concern in space engineering, and there have been extensive research activities aimed to address these issues.   approaches, tethering systems are quite effective to investigate the charging and magnetic field effects, and there are some recent review papers in the literature.

The TSS-1R is a shuttle-based system \cite{Dobrowolny} jointly developed and operated by NASA and the Italian Space Agency. It consists of a spherical satellite tethered by a space shuttle  with a 21.7 km of conducting tether with a total of resistance of 2100 ohms. The satellite traveled with a speed of  7800~m/s, roughly an order of magnitude faster than a sounding rocket payload. The satellite has higher potential than the ambient ionosphere, and electrons are collected onto the satellite, as the spacecraft charging process. An extra electron current is passed to the shuttle from the satellite. The tests  were performed in 1996,  however, the currents measured \cite{Thompson} were reported to be factors of two or three times greater than the predictions of the convectional Parker-Murphy (P-M) model and theory. This result was rather surprising since the P-M theory provides an upper limit for current collection. It is well known, that conventional sheath models are not complete without a treatment of the pre-sheath model that matches the sheath edge conditions to the ambient plasma.

On the scale of the TSS, the surrounding plasma is collisionless and the ambient electron gyro-radius is small; hence, it would seem that the TSS should deplete the immediate flux tube of electrons. If the tube was refilled along a length on which the plasma can be considered collisional, it is unclear whether the tube could carry a full thermal current to the sheath surface as assumed by the P-M theory. There are experimental data suggesting that the flux tubes indeed carry the full electron thermal flux to the sheath edge. Although many electron beam-emitting rockets have observed return currents in excess of the P-M limit, it was demonstrated by the  Cooperative High Altitude Rocket Gun Experiment (CHARGE) II experiment that this effect was due to the beam. Here, the electron-emitting mother payload experienced a return current in excess of the P-M limit, while the tethered daughter did not \cite{Mandell}. The Spectroscopy of Plasma Evolution from Astrophysical Radiation (SPEAR) I experiment also observed electron collection consistent with magnetic limiting in the high voltage sheath of a~positively-biased probe \cite{Katz}.

These sounding rocket experiments demonstrate that for even near stationary collections, the~ionosphere can refill the flux tube near the sheath. Understanding how this refilling occurs is an~important part of magnetic probe theory \cite{Laframboise}, and the ultimate source of the electrons constitutes the current closure issue \cite{Stenzel}. {In recent years, there are still investigations on the positive probe problems.  Cooke and Katz \cite{Cooke} argued that separating the issues of current closure and the current collection is empirically justified. They constructed a local probe model assuming an undisturbed plasma exists on the boundary of the iteration.  Their results are concise, and affected the NASA guidelines for spacecraft designs.

This chapter presents improvements on their previous work \cite{Cooke}, with more rigorous mathematical treatments on the pre-sheath model and  more measurements are included.

\section{Background for Positive Probe Sheath and Current}
The disturbed region of plasma surrounding a high voltage probe is commonly dissected into sheath and pre-sheath regions.  There are three basic sheath models, ``Orbit limited'', ``Space charge limited'', and ``Magnetic (P-M) limited''.  The last one is the model of our concerns due to the  Earth's strong magnetic field. In the ionosphere, electron gyration radii are on the order of centimeters, satellite~dimensions are on the order of meters, and collision mean free paths are on the order of  kilometers. As such, electrons move along the Earth's magnetic field lines, but remain essentially ``glued'' to the field lines  with tight spiral motions around them. Parker and Murphy  observed that for a potential distribution symmetrical about the magnetic field, the conservation of canonical angular momentum and energy yield a maximum cylindrical radius, $R_{PM}$, from  which an electron may reach the probe surface. This radius defines a flux tube from which electrons may be collected, and leads to a limiting current after specifying the current density in the flux tube. The P-M limit is weak because it can be violated by fluctuations \cite{Linson} or lack of symmetry \cite{Katz}. There are other popular theoretical  models  for spherically isotropic collections and collisionless conditions, in the absence of a magnetic field, but in the TSS-1R mission, the flow regime was never reached, and those other models are not discussed here. 

Whether the probe current is P-M or space-charge limited can be estimated by forming the ratio, between $R_{LB}$ and $R_{PM}$, where the former is the sheath by Langmuir-Blodgett. When the ratio is larger  than 1, the collection is P-M limited, and the space charge of electrons with $R_{PM}$ does not completely shield the probe. The probe potential therefore extends beyond $R_{PM}$, creating a region to which electrons are attracted but from which they cannot be collected. Even though these electrons are energetically capable of escaping, they cannot easily find an exit vector and form an $E \times B$ drifting shell of quasi-trapped electrons. Self-consistent modeling and laboratory experiments in support of the SPEAR-I  experiment demonstrate that this layer of circulating charges mostly overlies the P-M sheath region with negligible extension beyond $R_{PM}$. This contributes additional space charges and confines the sheath close to the PM radius.  % Figure \ref{Fig:illustration} illustrates this region. The actual trajectories of the quasi-trapped electrons reach from this shell to just above the probe surface.
\begin{figure}[ht]
%  \hspace{-2.cm}
  \centering
  \begin{minipage}[t]{.48\textwidth}
    \begin{center}
         \includegraphics[width=3.10in,height=2.4in]{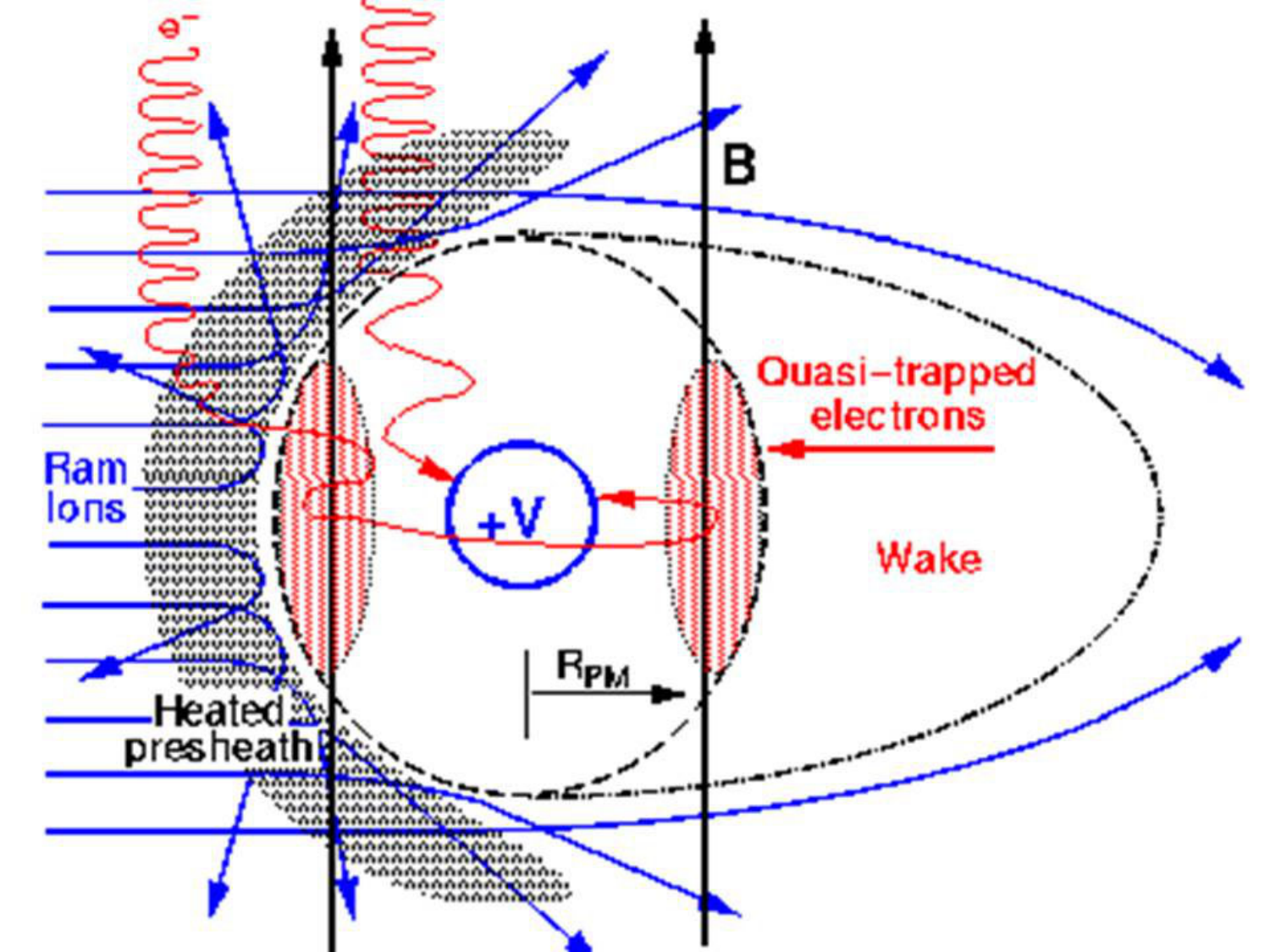}
    \end{center}
  \end{minipage}
    \begin{minipage}[t]{.48\textwidth}
    \begin{center}
      \includegraphics[width=3.0in,height=2.4in]{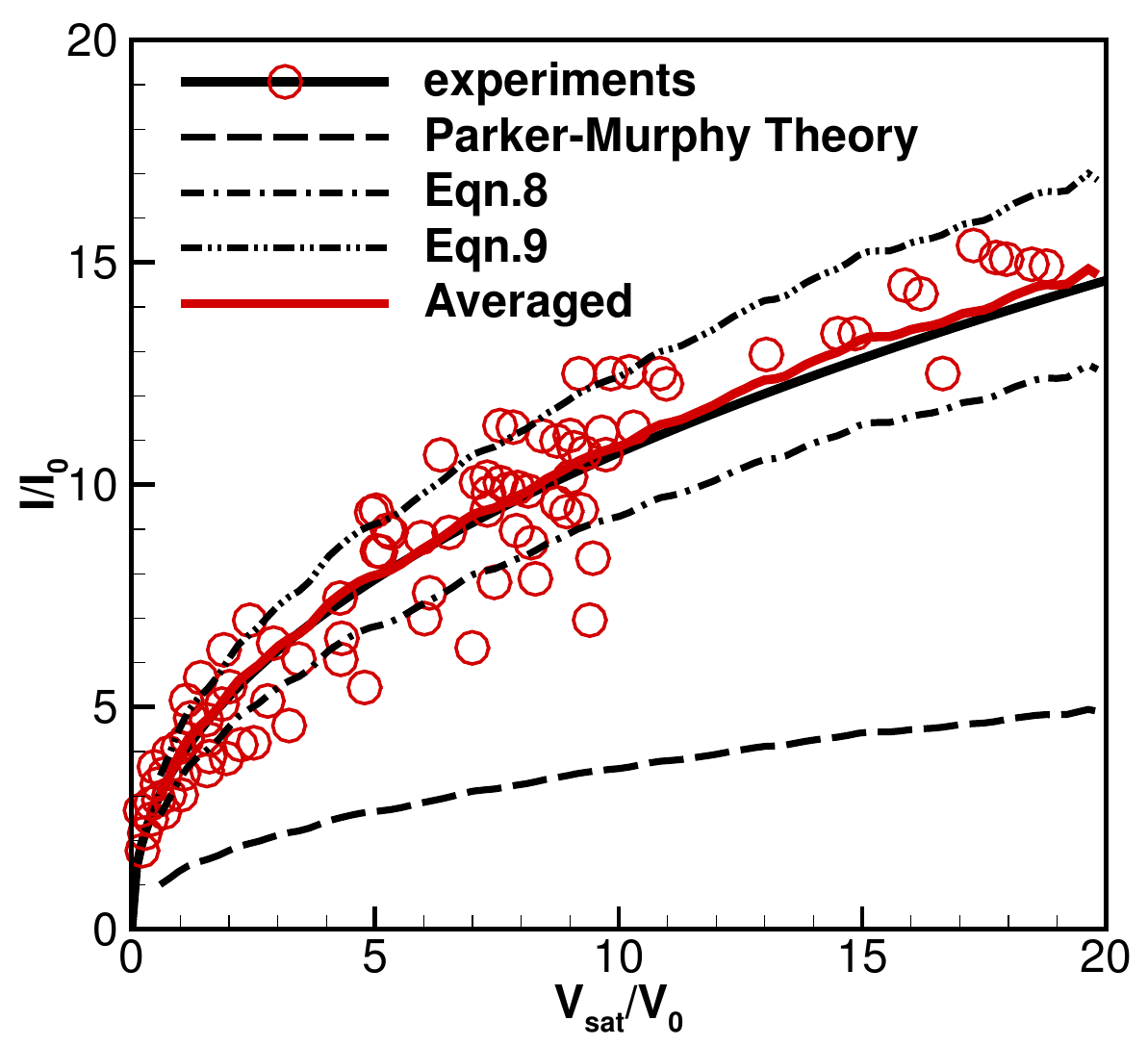}
    \end{center}
    \end{minipage}
     \caption{Top: schematic illustration of the interaction model\cite{Cooke}. Bottom: comparisons of current-voltage results. Symbols: measurements in space mission (solid~black line: the best fitting curve); Long dashed line: P-M theory;  Black dash-dot lines: new enhancement factors (solid red line: their average) }
    \label{Fig:illustration}
\end{figure}

\section{Different Approaches for Pre-Sheath Modeling}
One most important finding from the TSS-1R mission was that the Parker-Murphy theory significantly under-predicted the collected electron current. It was argued that the  sheath model is not enough to describe the problem, and the pre-sheath model must be considered as well. % Figure~\ref{Fig:presheath}~illustrates the relation across the pre-sheath, which is simplified as one-dimensional. 

Across the pre-sheath, behind the shock wave but in front of a sphere, there are two relations, one is the current and the other is the thermodynamic relation among heat flux, temperature, and current strength \cite{Cooke}:
\begin{equation}
  I  = \nabla V- \frac{1}{n} \nabla(nT) \approx \nabla V - \nabla T,   Q = -\frac{3}{2} T \nabla T + \frac{5}{2} T I,
 \label{eqn:current}
\end{equation}
where  $I$ is current density, $T$ is electron temperature, $V$ is plasma potential,  $n$ is electron or plasma density, and $Q$ is heat flux.  Equation (\ref{eqn:current}) assumes that the normalized electron number density gradient is negligible across the pre-sheath. 

The current continuity and heat relations are:
\begin{equation}
 \nabla \cdot  I  =0;   \nabla \cdot Q  = I^2.
 \label{eqn:dot}
\end{equation}

The first relation was discarded in their work due to the poor final agreements with measurements in the mission. From the other expression,  the following relation linking the temperature and potential changes is developed:
\begin{equation}
 \nabla^2 V  = \nabla^2 T,
\label{eqn:v2t2}
\end{equation}
a solution for the above relation is  $\nabla V = \nabla T -c_0$, where $c_0$ is a constant vector. The energy relations,  Equations (\ref{eqn:current}) and (\ref{eqn:dot}), transform into the following format:
\begin{equation}
     5 \left( \nabla T \right)^2 - \frac{9}{2}  \nabla T \cdot \nabla V + \left( \nabla V \right)^2  + \frac{3}{2} T \nabla^2 T =0.
 \label{eqn:gov}
\end{equation}
If the second order gradient term in Equation (\ref{eqn:gov}) is neglected, then there are two solutions:
\begin{equation}
    \nabla T = \frac{1}{2} \nabla V,  \nabla T = \frac{2}{5} \nabla V,
 \label{eqn:cooke_relation}
\end{equation}
 
%The flow is simplified as one-dimensional, and by using the following important mathematical integration relation:
%\begin{equation}
% \int_0^1 \sqrt{1+ax^2}dx  = \frac{1}{2} \sqrt{1+a} +  \frac{1}{2 \sqrt{a}}  \mbox{ln} ( \sqrt{a} + \sqrt{1+a})
%\end{equation}
The final general expression for the pre-sheath current due to thermal motion is obtained:
\begin{equation}
\begin{array}{rll}
 \frac{I_{PS}}{ 2 \pi R^2_{PM} I_0 } &=&  \frac{1}{2}  \big( 1+ (1+\chi)^{1/2} + \\
   && {\chi}^{-1/2} \mbox{ln}(\chi^{1/2} + (1+\chi)^{1/2} )  \big),
\end{array}
\label{eqn:aug1}
\end{equation}
where $\chi \equiv (T_s -T_0) /T_0$, $T_s$ and $T_0$ are the electron temperatures at the starting and ending of the pre-sheath. As the parameter $\chi$ increases, the current through the pre-sheath can be further simplified as \cite{Cooke}:
\begin{equation}
  I_{PS}  /( 2 \pi R^2_{PM} I_0)  =  \frac{1}{2} \big(1+ (1+\chi)^{1/2} \big).
\label{eqn:aug2}
\end{equation}

A new approach is  presented here. The above treatment can be improved and one of the major purposes of this paper is to report this new treatment.  One small defect in the above approach is that Equation (\ref{eqn:cooke_relation}) is inconsistent with Equation (\ref{eqn:v2t2}). With careful deviations, Equation (\ref{eqn:gov}) can transform to the following format if the higher order gradient term is reserved:
\begin{equation}
   \left( 7 \nabla T - 2\nabla V \right) \cdot \left( \nabla T - \nabla V \right) +  3 \nabla \cdot \left( T \nabla T \right) =0.
\end{equation}

At the start of the pre-sheath, there are two boundary conditions, $T(0) =T_0$, and $V(0) =0$. If the probe potential is high enough, then the potential at the sheath edge takes specific values, $V(x=L) =V_s  = 0.493 $ eV \cite{Parrot}. The temperature at the sheath edge is to be determined.

It can be shown, if the current is zero, $I=0$, then there are two specific solutions to the potential and temperature:
\begin{equation}
\begin{array}{rll}
  V(x)  &=&  \sqrt{\frac{V_s (V_s+2T_0) }{L} x +T_0^2} -T_0; \\
  T(x)  &=& \sqrt{\frac{V_s (V_s+2T_0) }{L} x +T_0^2},
\end{array}
 \label{eqn:quadratic}
\end{equation}
and Equation (\ref{eqn:v2t2}) holds.  The above solutions imply that there is no current across the pre-sheath, based~on Equation (\ref{eqn:current}).

Equations (\ref{eqn:v2t2}) and  (\ref{eqn:gov}) can be simplified as two ordinary differential equations for a one-dimensional situation, and further can be merged into an ordinary Abel differential equation of the second kind.  With proper boundary conditions, there are exact mathematical solutions, however, the solutions are rather complex for use and are not adopted in this work. The following simpler approach is developed instead, and they transform the ordinary differential equation into an algebraic equation.

First the \emph{X}-coordinate for Equations (\ref{eqn:v2t2}) and (\ref{eqn:gov}) is normalized by using the pre-sheath length, \mbox{and the} domain of interest transforms into $0 \le X \le 1$.  A new variable for the temperature is defined as $t(X)= T(X)- T_0$, with the boundary values $t(X=0) =0$ and  $t(X=1) =t_s=T_s-T_0$, \mbox{where the} latter is the property of our interest. Then, Equation (\ref{eqn:v2t2}) transforms as:
\begin{equation}
   \nabla^2 v(X) =  \nabla^2  t(X),
\end{equation}
which has a general solution $v(X) = t(X) +c_0 X + c_1 =  t(X) +c_0 X$.  With these transformations, Equation (\ref{eqn:gov}) changes to:
\begin{equation}
    5 ( \nabla t)^2  - \frac{9}{2} \nabla t \cdot \nabla v  +  (\nabla v)^2   + \frac{3}{2}  \left( t+ T_0 \right)   \nabla^2 t=0
\end{equation}
This equation can further transform if we introduce one new variable,  $\alpha \equiv t_s /v_s >0$, with $c_0  = v_s -t_s = (1-\alpha)v_s$, then
\begin{equation}
 \left(   \frac{3}{2}  \nabla t - \frac{1 - \alpha}{\alpha} t_s  \right)     \left( \nabla t - \frac{1 - \alpha}{\alpha} t_s  \right)    + \frac{3}{2}  \left( t+ T_0 \right)  \nabla^2 t =0
\label{eqn:ode}
\end{equation}
Across the pre-sheath, $0 \le \alpha \le \infty$, its value is bounded.

Enlightened  by the solution format developed by Cooke and Katz \cite{Cooke}, the following general power function format for $t(X)$ is assumed, with which the above ordinary differential equation can degenerate to an algebraic equation:
\begin{equation}
    t(X)  = t_s X^n,
\label{eqn:deft}
\end{equation}
where $n$ is a free parameter, limited between 0 and 1,  which mimics the quadratic formats in Equations~(\ref{eqn:cooke_relation}) and (\ref{eqn:quadratic}). At the sheath ending edge, with $X=1$, Equation (\ref{eqn:ode}) simplifies as:
\begin{equation}
   \frac{1}{\alpha^2}  - \left(\frac{5}{2}n+2 \right)  \frac{1}{\alpha}  + \frac{3}{2} n (n-1) \frac{1}{\alpha} \frac{T_0}{v_s}  + 3n^2 +n+1 =0.
\label{eqn:alpha}
\end{equation}
where parameters $T_0 =0.13$ ev, $v_s=0.493$ V are used in the above equation. These two parameters are the reference values for the TSS-1R mission performed in the ionosphere.
If $n>1$, there is no real solution for the above equation. This fact is compatible with our assumptions on the scope for the parameter range.  There are two real roots of $\alpha$ for the above equation.

Figure \ref{Fig:illustration} shows additional experimental results collected in the TSSR-1R mission with different free stream conditions. In the literature, a group of averaged free stream parameters was adopted, and computed a smooth curve for the Parker-Murphy theory: 
\begin{equation}
%\begin{array}{rll}
  \frac{I}{I_0} = \frac{1}{2} +   \sqrt{ \frac{V}{V_0} },  I_0 =  \pi a^2  e n \sqrt{ \frac{8kT_e}{\pi m_e} };   V_0 =  \frac{B^2 a^2 e}{2m_e},
%\end{array}
\end{equation}
where $V$ is the measured voltage, $B =0.32$ Gauss is the Earth's magnetic field strength, $a$ is the radius for the spherical satellite, $e$ is the electron change, $n$ is the ambient electron density, $T_e$ and $m_e$ are the electron temperature and mass. That curve is included in this figure as a long dashed line at the bottom. Space experimental measurements collected during the  TSS-1R mission are represented by circles, and the best fitted curve is presented as the solid black line.  These data are very valuable because they reflect plasma behaviors in the ionosphere, and can not be created in a ground environment. \mbox{It is} evident that there are large scatters because the data are collected within different time periods, and with slightly different free stream electron temperatures $T_0$,  and different values for number density $n$. Two current enhancement curves are computed and  included in this figure as two dash-dotted lines, the computations are based on Equations (\ref{eqn:aug1}) and (\ref{eqn:aug2}), the specific values are computed by averaging the enhancement values within the range of $0.5<n<1.0$. The large scatters in the experiments indicate it is impossible to use one curve to describe these data. However,  these two new curves for current enhancement coefficient bound most measurements, and successfully captured the trends in the experimental data. They are more accurate than the traditional P-M theory.  The past work \cite{Cooke} adopted a specific enhancement factor with one curve, and included only seven measurement data points available at the time when the paper was published. The single curve represents the seven measurement points very well, more experiments are available now, and they indicate that one curve cannot represent the results very well. Figure \ref{Fig:illustration} also includes  a solid red line which is computed by averaging these two new lines.  Evidently, the averaged curve agrees quite well with the best curve-fitting results for the  experimental measurements.  

\section{Conclusions}
\label{sec:conclusion}
Improvements on analyzing current enhancements during the TSS-1R mission is reviewed. A more rigorous mathematical treatment procedure is developed and presented. Compared with the past treatment, the second order differential term in the governing partial differential equation is preserved for the pre-sheath model. By assuming special formats for the temperature profiles within the pre-sheath, a family of general, exact solutions to the partial differential equation are obtained. The results indicate that a smaller $\alpha$ value is more reasonable. The~$n$~values in this new model do not lead to appreciable differences in the predicted current enhancement results. The work also confirms the past treatment is a special case with $n=1$, and the past work is reasonable and solid.    

In the past effort, seven data points were included for comparison. The data agree well with their formula, and predicted a current  enhancement coefficient of 2.5.   This study includes more experimental measurements and recovers a new curve.  The large scatters in the measured  TSS-1R space mission indicate that it is challenging to describe the measurements by using one curve. Instead, the two curves developed in this study properly bound the measurements. The curve computed by averaging these two curves agrees surprisingly well with the best curve-fitting  results for the experimental measurements.  The averaged current enhancement coefficient is estimated as 2.95, which is slightly larger than 2.5, as recommended by Cooke and Katz \cite{Cooke}.

\vspace{12pt}

\end{document}